\documentclass[preprint,authoryear,12pt]{elsarticle}
\usepackage{graphicx}
\usepackage{amsmath,amssymb}

\newcommand\calE{{\cal E}}
\newcommand\bfcalE{\boldsymbol{\cal E}}

\newcommand{\Rsun}{R_\odot}

\journal{Physics of the Earth and Planetary Interiors}

\begin{document} 

\begin{frontmatter}



\title{Solar Magnetic Fields}


\author[1]{Alan W Hood}
\ead{alan@mcs.st-and.ac.uk}
\author[2]{David W Hughes}

\address[1]{School of Mathematics and Statistics,\\
University of St Andrews,\\
KY16 9SS, UK}
\address[2]{Department of Applied Mathematics,\\
University of Leeds,\\
LS2 9JT, UK}

\begin{abstract}
This review provides an introduction to the generation and evolution of the Sun's magnetic field, summarising both observational evidence and theoretical models. The eleven year solar cycle, which is well known from a variety of observed quantities, strongly supports the idea of a large-scale solar dynamo. Current theoretical ideas on the location and mechanism of this dynamo are presented. 

The solar cycle influences the behaviour of the global coronal magnetic field and it is the eruptions of this field that can impact on the Earth's environment. These global coronal variations can be modelled to a surprising degree of accuracy. Recent high resolution observations of the Sun's magnetic field in quiet regions, away from sunspots, show that there is a continual evolution of a small-scale magnetic field, presumably produced by small-scale dynamo action in the solar interior.

Sunspots, a natural consequence of the large-scale dynamo, emerge, evolve and disperse over a period of several days. Numerical simulations can help to determine the physical processes governing the emergence of sunspots. We discuss the interaction of these emerging fields with the pre-existing coronal field, resulting in a variety of dynamic phenomena.
\end{abstract}

\begin{keyword}
Solar magnetic fields \sep Solar cycle \sep Dynamo  \sep Flux emergence \sep Global coronal magnetic field


\end{keyword}

\end{frontmatter}

% \linenumbers

\section{Introduction}\label{sec:introduction}
The Sun's magnetic field is generated through dynamo action in the solar interior. While the key
physical processes involved with this dynamo are hidden from view, the consequences of the
resulting magnetic field can be seen in the solar atmosphere. The large-scale magnetic field is 
characterised by the appearance of sunspots, regions of intense magnetic field, the number of which varies with a cycle of approximately eleven years. In addition, there is a
small-scale magnetic field that varies on a much shorter timescale. The appearance of sunspots
and the continual motion of the small-scale field ensure that the magnetic field of the solar atmosphere is very dynamic, with many solar phenomena owing their existence to this evolving field. The aim of this article is to describe the present theoretical ideas behind various aspects of the solar magnetic field. In particular, we shall look at current ideas for the dynamo mechanism(s) responsible for the generation of large- and small-scale fields, and will then consider the dynamical behaviour of the surface manifestations of these fields, how they might burst through the solar surface and interact with magnetic field in the solar atmosphere. We have tried to address an audience familiar with aspects of the Earth's magnetic field, but maybe less so with those of the Sun. Thus it is perhaps appropriate first to present a short description of the basic properties of the solar interior and the solar atmosphere.

\subsection{Properties of the solar interior}
The radius of the Sun is $\Rsun = 696$\,Mm, or equivalently $69\,600$\,km. From the theory of stellar structure, it follows that the interior of the Sun has three physically distinct regions. At the centre, extending to approximately $0.25 \Rsun$, is the core, in which nuclear reactions fuse hydrogen into helium, with the associated release of energy. From $0.25 \Rsun$ to $0.71 \Rsun$, through what is known as the radiative zone, energy is transferred by the emission and absorption of photons. The outer $30\%$ by radius of the Sun (which, however, constitutes only $2\%$ of the solar mass) is the convection zone; here, the increase in opacity leads to a temperature gradient sufficiently steep to trigger convective motions. Recently, in only the past thirty years or so, it has become possible actually to determine properties of the solar interior via observation and inversion of the sound waves propagating through the interior --- a field known as \textit{helioseismology}. Of particular interest for theories concerning the generation of the solar magnetic field is the deduction of the solar interior rotation rate, $\Omega$. Although, through observations of surface magnetic features, it has been known for hundreds of years that the solar surface rotates differentially, with the equator rotating faster than the poles, with periods of approximately 25 and 34 days respectively, the nature of the internal rotation rate remained a matter for theoretical speculation. The actual internal rotation rate revealed by helioseismology turned out to be rather surprising \citep{Schou_etal_98}. The differential rotation in latitude of the surface is maintained throughout the convection zone, with very little radial variation; the radiative zone is in solid body rotation, with an angular velocity comparable with that of the surface at approximately $30^\circ$ latitude. The abrupt change in angular rotation between the convection and radiative zones is then accommodated in a very thin shear layer, known as the \textit{tachocline}; $\partial \Omega / \partial r > 0$ at low latitudes, $\partial \Omega / \partial r < 0$ at higher latitudes. Although very thin in extent, occupying just a few per cent of the solar radius, the existence of the tachocline is of great significance both in determining the long-term evolution of the Sun and also for its possible role in the generation of the solar magnetic field; all currently known aspects of tachocline dynamics are covered in the recent volume edited by \cite{HRW_07}.

Magnetic field pervades all regions of the Sun. In the radiative zone, it is presumably responsible for enforcing the observed solid body rotation \citep[see][]{MW_87} and hence the dynamical behaviour of the magnetic field is a key factor in the evolution of the radiative zone. For the purposes of this review however, which concentrates on the observed solar magnetic field and the processes responsible for that field, we shall confine our attention to the magnetic field in the convection zone (including the tachocline) and the solar atmosphere. A comprehensive discussion of all aspects of stellar magnetic fields can be found in \citet{Mestel_99}.

\subsection{Properties of the solar atmosphere}
The visible \lq surface\rq\ of the Sun is the \textit{photosphere}. The temperature is around $6\,000$\,K and the radiation is in the form of 
white light. The thickness of the photosphere is only 500\,km; the base of the photosphere is where the optical depth is $\tau_{500} = 1$ 
and the top is where the temperature reaches its minimum value.

The most obvious surface features are \textit{sunspots}; an example is shown in Figure~\ref{fig:sunspot}. Sunspots have been observed in detail since the invention of the telescope in the 17th century; they were first identified as regions of extremely strong magnetic field by Hale (1908). The typical magnitude of the magnetic field in a sunspot is the order of $3\,000$\,G or $0.3$\,Tesla. The magnetic field manifested in sunspots is generated by dynamo action in the solar interior.

High resolution images of the photosphere, away from sunspots, show evidence of convection through \textit{granules} and \textit{supergranules}. Granules are approximately 1000\,km in horizontal extent and cover the entire surface of the Sun. Hot plasma rises up from the solar interior in the bright centres, spreads out across the surface and then cools and sinks into the interior along dark lanes. The lifetime of an individual granule is about 20 minutes, with the granulation pattern continually changing as newly formed granules push aside old ones. The horizontal plasma flow within a granule can reach speeds of around 7\,km\,s$^{-1}$.  
 
Supergranules are much larger, with a size of about 14\,000\,km. They are not visible in white light but are revealed through Doppler velocity measurements. Like the granules, which lie inside the supergranules, they cover the entire photosphere and are  continually changing. Individual supergranules last for around 24 hours and have much slower plasma speeds of about 0.5\,km\,s$^{-1}$. 

Above the photosphere, the \textit{chromosphere} is an irregular layer with a thickness of around 10\,000\,km, in which the temperature rises from $6\,000$\,K to about $20\,000$\,K. The plasma now emits in the wavelength of Hydrogen alpha ($H\alpha$); features seen in $H\alpha$ are thought to outline the local magnetic field lines. At the top of the chromosphere, the temperature rises sharply through a thin layer called the \textit{transition region}.

The \textit{corona} is the Sun's outer atmosphere. It is visible from Earth during total eclipses of the Sun as a diffuse white region surrounding the Sun. The temperature in the corona increases dramatically, to the order of 1 million~K, and the wavelength of the plasma emission is in the X-ray and Extreme Ultra-Violet ranges. One major problem in solar physics is to explain why the corona is so hot. The corona displays a variety of features that are entirely controlled by the local, evolving coronal magnetic field, including streamers, plumes, and loops. 

\subsection{Modelling philosophy}
To reproduce through numerical simulations all the detailed observations of the evolution of the Sun's magnetic field, obtained from various instruments, is certainly currently impossible since there are simply too many distinct timescales and lengthscales to resolve in numerical simulations. Indeed, even with the continuing increase in computational power and affordability, such direct modelling will remain impossible for the foreseeable future. There are many physical and atomic processes required for a complete modelling of the solar atmosphere. For example, computationally demanding processes include (i) radiative transfer effects in the photosphere and chromosphere and (ii) optically thin radiative losses, thermal conduction and heating in the corona. In addition, in the cool photosphere and low chromosphere, partial ionisation must be considered. All of these processes require knowledge of, for example, the ion abundances, whether the plasma is in local thermodynamic equilibrium or not, and whether the plasma is in ionisation balance.

Controlled numerical experiments allow one to specify which physical effects are to be studied. Hence, the modelling philosophy is to consider a simpler initial state, for example one that is in equilibrium, and then modify it in a controlled manner to see how one effect at a time influences the evolution. Incorporating too many variations all at once obscures the important or dominant physical processes.

\subsection{Outline}
The outline of the paper is as follows. Section~\ref{sec:MHDequations} presents the MHD equations used in modelling the solar dynamo, the Earth's dynamo and the magnetic field evolution in the solar corona. In Section~\ref{sec:cycle} the observational evidence for the solar cycle is presented. Next, Section~\ref{sec:dynamo} describes the current state of modelling the generation of the Sun's magnetic field, both the large-scale magnetic field, on the scale of active regions, and the small-scale field, on the scale of the granulation. Section~\ref{sec:comparison} compares the dynamo mechanisms of the Sun and the Earth, and looks at the similarities and differences in the computational models of both types of dynamo.  Moving to the atmospheric consequences of the solar dynamo, Section~\ref{sec:coronalB} considers the nature of the coronal magnetic field, and the manner in which it is influenced by the movement of the photospheric field. Returning to the formation of individual active regions, Section~\ref{sec:emergence} investigates how the strong magnetic field generated in the solar interior actually emerges through the photosphere and rises into the corona, where it interacts with the pre-existing field. During this process, many of the observational phenomena, whose properties are linked to the solar cycle, are seen to occur in a self-consistent manner. The final section summarises the paper.

\section{MHD Equations}\label{sec:MHDequations}
The gas of the Sun, both in the interior and the atmosphere, is ionised. Its dynamical
evolution can be described by the equations of MagnetoHydroDynamics (MHD), a set of coupled, non-linear partial differential equations, which may be stated as
\begin{equation}
\rho \left( \frac{\partial {\bf u}}{\partial t} +({\bf u}\cdot \nabla){\bf u}+ 2 {\bf \Omega}\times {\bf u} \right) = -\nabla p + {\bf j}\times {\bf B} + \rho{\bf g} + {\bf F}_{\nu},
\label{motion}
\end{equation}
\begin{equation}
\frac{\partial\rho}{\partial t} + \nabla \cdot (\rho {\bf u}) = 0,
\label{continuity}
\end{equation}
\begin{equation}
\frac{\partial {\bf B}}{\partial t} = \nabla \times ({\bf u}\times {\bf B}) - \nabla \times (\eta \nabla \times {\bf B}),
\label{induction}
\end{equation}
\begin{equation}
\frac{\rho^\gamma}{\gamma -1} \frac{\partial}{\partial t}\left (\frac{p}{\rho^\gamma}\right ) +\frac{\rho^\gamma}{\gamma -1} ({\bf u}\cdot \nabla )\left (\frac{p}{\rho^\gamma}\right ) =  \nabla \cdot (\kappa \nabla T) + \frac{j^2}{\sigma} + Q_{visc},
\label{energy}
\end{equation}
\begin{equation}
{\bf j} =\frac{1}{\mu}\nabla \times {\bf B},\qquad \nabla \cdot {\bf B} = 0,
\label{ampere}
\end{equation}
\begin{equation}
p=\frac{\rho}{\tilde{\mu}}RT.
\label{gaslaw}
\end{equation}
Here $\rho$ is the mass density, ${\bf u}$ the plasma velocity, ${\bf \Omega}$ the rotation vector, $p$ the plasma pressure, ${\bf j}$ the current density, ${\bf B}$ the magnetic induction (though commonly called the magnetic field) and $T$ the temperature; ${\bf g}$ is the gravitational acceleration, $\gamma$ the ratio of specific heats, commonly assumed to be $5/3$, $\tilde{\mu}$ is the mean molecular weight, $R$ is the gas constant, $\mu$ the magnetic permeability, and $\eta$ the magnetic diffusivity, with $\eta = 1/\mu \sigma$ and $\sigma$ the electrical conductivity. In the momentum equation (\ref{motion}), ${\bf F}_\nu$ represents the viscous force for a Newtonian fluid. The thermal conductivity tensor, $\kappa$, is normally taken as isotropic in the solar interior but is highly anisotropic in the solar corona. Ohm's law in its simplest form has been used to derive the induction equation. The energy equation consists of the adiabatic terms together with sources of heating through Ohmic and viscous heating, including shock heating. However, additional terms such as optically thin radiation, radiative transfer effects, partial ionisation, etc.\ may need to be included in different levels of the solar atmosphere. These equations can be solved (at least in certain parameter regimes) using 3D 
MHD simulation codes on a parallel computer. For example, compressible magnetoconvection has been modelled by \cite{Matthews_etal_95}, magnetic events in the solar atmosphere by the Lare3d code (Arber et al., 2001).  

These basic equations are also the appropriate fundamental equations for modelling the Earth's magnetic field, although it should be noted that sound waves have no dynamic effects in the Earth, and hence it is then advantageous to adopt equations in which they are neglected. Consequently, geodynamo models typically adopt either the Boussinesq approximation (in which sound waves and stratification are ignored) or, more realistically, the anelastic approximation (in which sound waves are again filtered out, but stratification retained). Indeed, the anelastic approximation may also be valid for the deep solar interior, and global solar dynamo models often employ this approximation \citep[e.g.][]{Clune_etal_99}. Under the Boussinesq approximation, density variations appear only in the buoyancy term in equation~(\ref{motion}), and equation~(\ref{continuity}) simplifies to $\nabla \cdot {\bf u} = 0$; viscous and Ohmic heating are omitted from equation~(\ref{energy}). The anelastic approximation, which is less restrictive (but inevitably more complicated) than the Boussinesq approximation, considers departures from an adiabatic reference state, leading to changes in equations~(\ref{motion}), (\ref{continuity}) (which becomes $\nabla \cdot ( {\bar \rho} {\bf u} ) = 0$) and (\ref{energy}). A detailed discussion of the two approximations may be found in \cite{Lantz_Fan_99}.

The relative importance of the various terms in the MHD equations can be gauged by the size of various dimensionless parameters. The ratio of advective to diffusive terms in the momentum equation~(\ref{motion}) and the induction equation~(\ref{induction}) are given, respectively, by the Reynolds number $Re$ and the magnetic Reynolds number $Rm$, defined by
\begin{displaymath}
Re = \frac{UL}{\nu}, \qquad
Rm = \frac{UL}{\eta},
\end{displaymath}
where $U$ and $L$ are characteristic velocity and length scales and $\nu$ is the kinematic viscosity. Both Reynolds numbers are large in the solar interior ($Re = O(10^9)$, $Rm = O(10^6)$), with their ratio $Pm = Rm/Re = \nu / \eta$, the magnetic Prandtl number, being small. In the solar corona, $Rm = O(10^8)$ --- $O(10^{12})$ while $Re$ is also larger than unity and is of probably $O(10^{4})$. The value of $Re$ is based on the kinematic viscosity 
parallel to the magnetic field. Large values of $Re$ are characteristic of turbulent flows; large values
of $Rm$ mean that resistive effects are restricted to thin regions where the current density is large. However,
if the plasma flows are turbulent, then $Rm$ can be reduced significantly. 
The regular Prandtl number is defined by $Pr=\nu/\kappa$. The importance of the viscous to Coriolis terms in (\ref{motion}) is given by the Ekman number, defined by
\begin{displaymath}
E = \frac{\nu L^2}{2 \Omega} .
\end{displaymath}
Alternatively, the relative importance of these two terms is expressed, conversely, as the Taylor number, defined by $Ta = E^{-2}$.

\section{Solar Cycle Variations}\label{sec:cycle}

\begin{figure}
\centering
\includegraphics[width=0.65\textwidth]{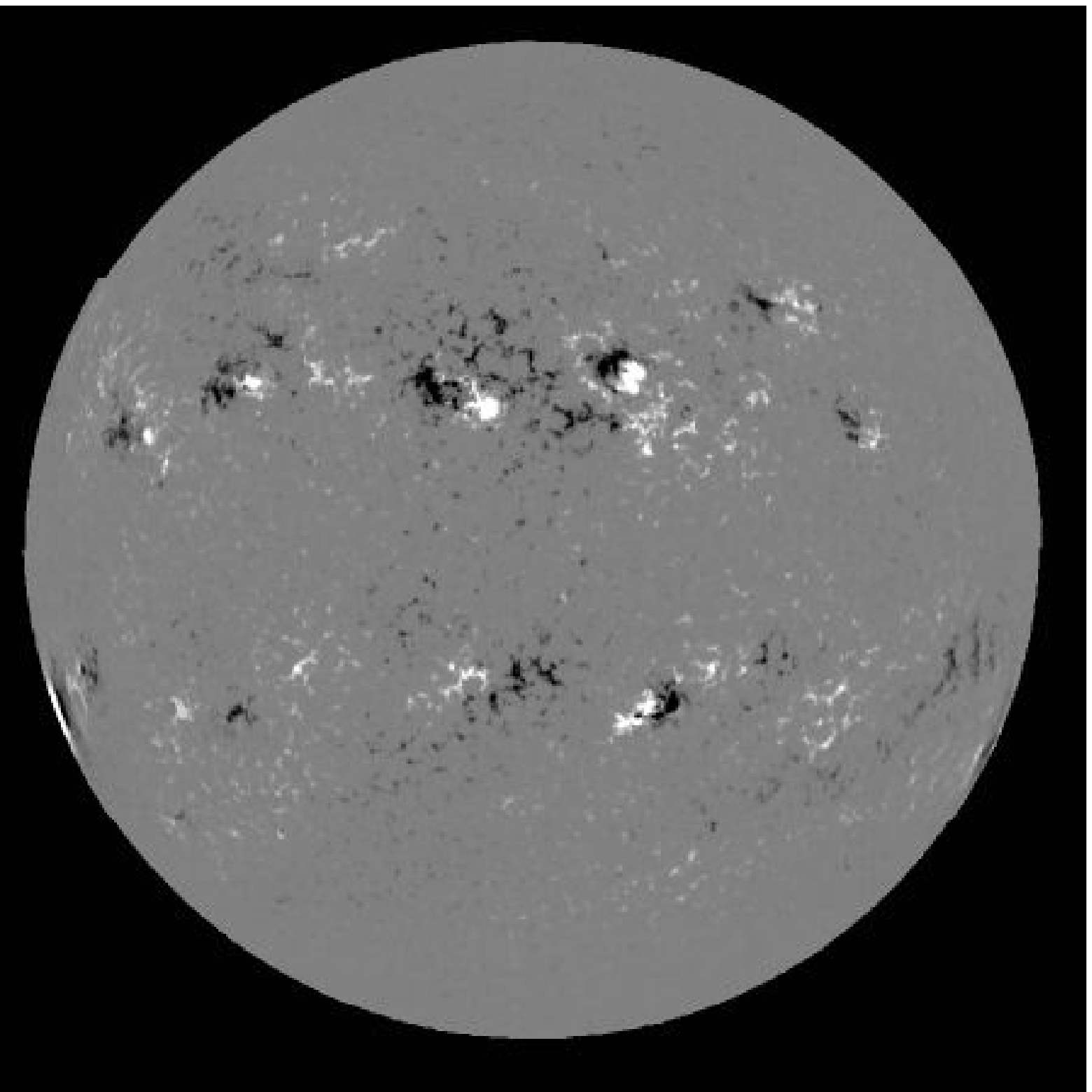}
\includegraphics[width=0.65\textwidth]{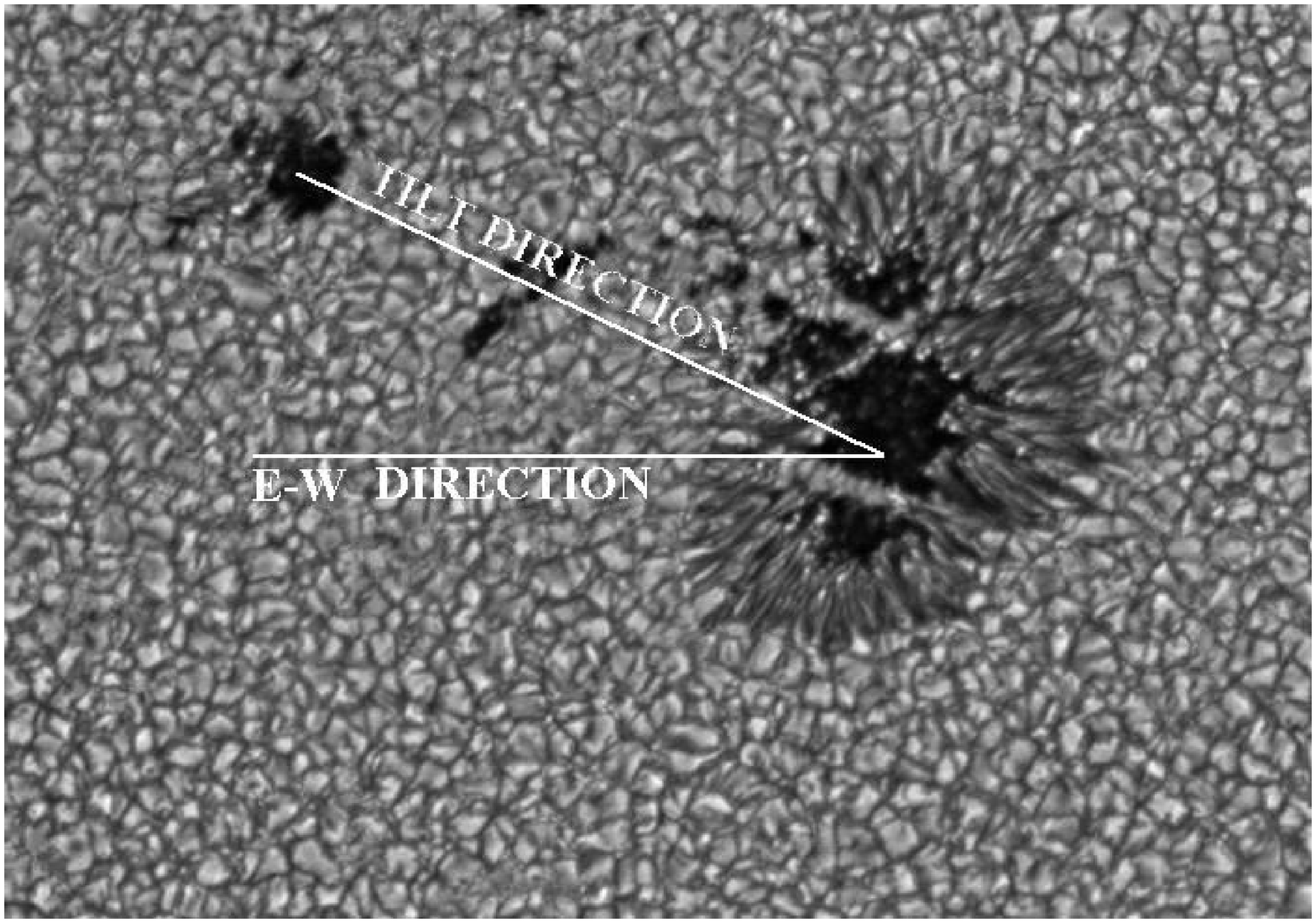}
\caption{The top figure is a magnetogram showing the line of sight magnetic field 
(from bison.ph.bham.ac.uk/\~wjc/Teaching/joys\_law.jpg). The leading (right) and trailing (left) polarities have the same orientation in the northern hemisphere and the opposite orientation in the southern hemisphere. The bottom figure shows a white light image of a sunspot pair 
(from www.hao.ucar.edu/research/siv/images/hale\_rule.gif) and clearly illustrates the tilt with respect to the East-West direction.}
\label{fig:sunspot}
\end{figure}
The main evidence for the solar cycle comes from studying the evolution of sunspots over time. 
The number of sunspots and the area they occupy vary with a rough period of 11 years. At the start of a cycle, these quantities are low. They rapidly increase in value to reach a maximum before decreasing, more slowly this time, down to minimum values. The existence of the Sun's cyclic variation is evident from observations of (i) 10.7 cm solar flux, (ii) the total solar irradiance, (iii) solar flares and Coronal Mass Ejections (CMEs), (iv) interactions with the geomagnetic field, (v) cosmic rays at the Earth, and (vi) radioisotopes in tree rings and ice cores.

Measurements of isotopes such as $^{14}C$, found in trees, and $^{10}Be$, deposited in ice cores, reflect the number of cosmic rays reaching the Earth; they provide a proxy measure of the solar magnetic field, since a strong field emanating from the Sun acts to shield the Earth from cosmic rays. Records of such radioisotopes provide information on the 11-year solar cycle for the past 600 years, and on longer time variations of the magnetic field stretching back nearly 10\,000 years \citep{Beer_2000}. They show that solar activity is modulated in an aperiodic and apparently chaotic manner, with episodes both of enhanced and diminished magnetic activity \citep{Weiss_2010}. The most recent dramatic example of the latter was the \textit{Maunder minimum}, the period from 1645-1715, during which very few sunspots appeared.

\subsection{Sunspot cycle}\label{sec:sc}
Sunspots typically appear as a bipolar region (an active region), with a dominant sunspot of one 
polarity leading (in the sense of the solar rotation) the trailing, opposite polarity region. At the start of a new sunspot cycle, the sunspots appear at high latitudes, around $\pm 30^\circ$, and, as the cycle continues, the new sunspots emerge closer and closer to the equator. The trailing polarity may be another sunspot or it can be a more diffuse magnetic region. Hale's polarity law states that at any given time, the polarity of the leading (and hence also the trailing) spot is the same in a given hemisphere, but is reversed from northern to southern hemispheres. This is illustrated in Figure~\ref{fig:sunspot} (top) which shows a magnetogram, with white corresponding to positive polarity and black to negative; in the northern hemisphere, the leading polarity is positive and in the southern hemisphere it is negative. During the next solar cycle, this reverses and the leading polarity in the northern hemisphere will now be negative (so a full magnetic cycle takes approximately $22$ years). 

In addition, Joy's law states that, while the bipolar region is essentially in the East-West direction, 
the leading polarity is always slightly closer to the equator than the trailing polarity. Figure~\ref{fig:sunspot} (bottom) shows a bipolar sunspot pair in the northern hemisphere and the corresponding tilt angle. These properties are important and must be built into models that use the photospheric magnetic field to extrapolate the field into the corona. The magnetic field strength in a new sunspot grows in magnitude over a few days. Most sunspots decay within a few days after that, although extremely large sunspots and active regions can last for up to 28 days. When the sunspot starts to decay, supergranulation eventually assists in the breakup of the strong magnetic field region, a process that can be modelled by a diffusion equation.

\subsection{Small-scale fields}
\begin{figure}
\centering
\includegraphics[width=0.8\textwidth]{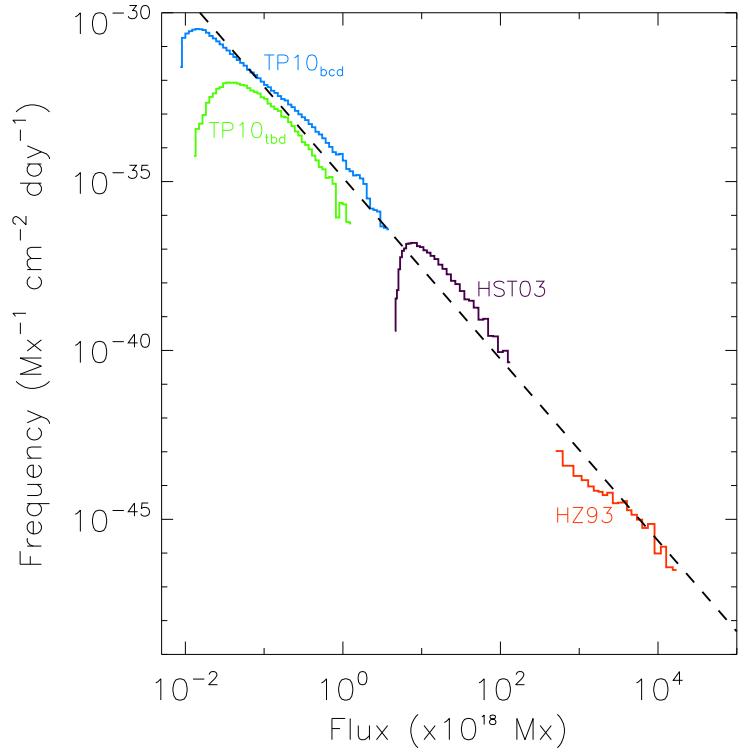}
\caption{Log-log plot of the frequency of emergence against emergence event flux. The fit (dashed line) is a power-law distribution with an index of $-2.7$. Distributions plotted are from Kitt Peak (HZ93 -- Harvey and Zwaan, 1993), MDI (HST03 -- Hagenaar, Schrijver and Title, 2003) and SOT (TP10$_{bcd}$/TP10$_{tbd}$ -- Thornton and Parnell, 2011). (Based on Thornton and Parnell, 2011.)}
\label{fluxhistogram}
\end{figure}
Sunspots and active regions are the most obvious indicators of the solar dynamo. However, away from the active regions, there is a 
continual emergence and evolution, including fragmentation, coalescence and cancellation, of much smaller-scale fields. Such fields 
emerge as \textit{ephemeral regions} within supergranules and on much smaller scales within granules. The recent observations from 
the Solar Optical Telescope on board the Hinode mission has the resolution to investigate the distribution of magnetic fields, from very 
weak fields with a flux of around $10^{16}$Mx to the large active region sunspots with a flux of up to $10^{22}$Mx. How important is this 
small-scale field? Figure~\ref{fluxhistogram} shows the probability distribution function of emergence events  in units of Mx$^{-1}$\,cm$^
{-2}$\,day$^{-1}$ as a function of the magnetic flux. The key point is that the various data sets can be approximated by a single power 
law fit with a slope of $-2.7$. Since the slope is greater than 2, this implies that there is more photospheric magnetic flux contained within 
the small-scale field than in the active regions. A single power law indicates that there is no preferred scale for the emergence of
magnetic flux at the solar photosphere, providing useful constraints on solar dynamo models.

\begin{figure}
\centering
\includegraphics[width=0.7\textwidth]{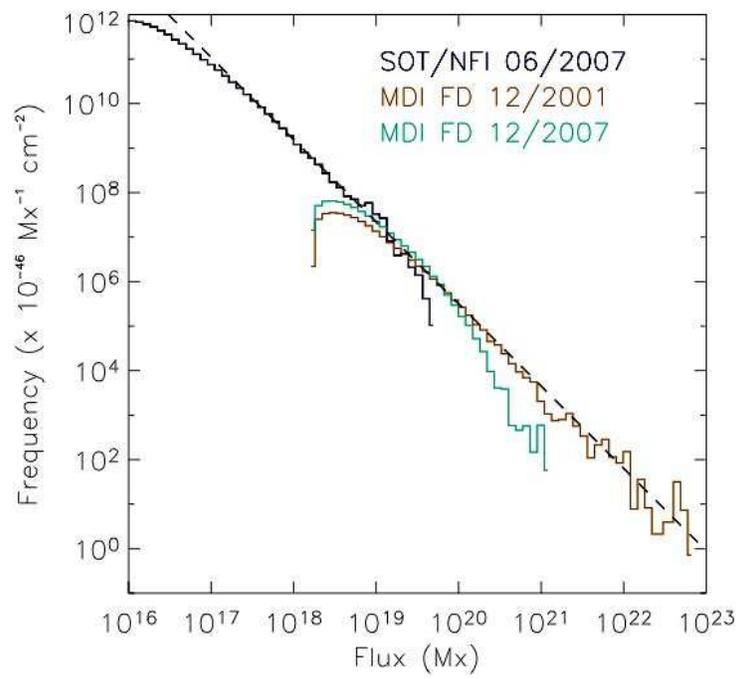}
\caption{Log-log plot of the frequency of occurrence of features against feature flux. The fit (dashed line) has a slope of $-1.85$. (From Parnell et al., 2009.)}
\label{parnellfig5b}
\end{figure}
Figure~\ref{parnellfig5b} shows the probability distribution function of all magnetic feature fluxes at an instant in time, in units of Mx$^{-1}$\,cm$^{-2}$ as a function of magnetic flux. The results from the SoHO mission show the distribution function at solar minimum (2007) and solar maximum (2001). The results from SOT on Hinode, which can observe the small-scale field more accurately, are also shown. It is clear that at solar maximum the tail of the distribution function is enhanced and a single power law fit is possible. At solar minimum there is now less large-scale flux as there are few active regions. However, the small-scale flux appears to maintain the same slope at both stages of the solar cycle. Thus, while the large-scale magnetic fields in sunspots varies during the solar cycle, there is no decrease in the small-scale field during solar minimum.

The small-scale magnetic features are, therefore, extremely important. These are moved around the solar photospheric surface by the continually changing granulation and
supergranulation flow patterns; the movement of these magnetic features is often referred to as the \textit{magnetic carpet}. A simple estimate of the Poynting flux, the magnetic energy flowing into the upper solar atmosphere due to the motion of the magnetic carpet, indicates that there is sufficient energy to heat the solar corona and explain why the coronal temperature is around a million degrees (Mackay et al., 2011).

\section{The Solar Dynamo}\label{sec:dynamo}

It is almost universally accepted that the solar magnetic field is the result of some sort of hydromagnetic dynamo action, in which plasma motions maintain the magnetic field against its tendency otherwise to decay. The alternative possibility, namely that the observed field is derived from a very slowly decaying primordial field, is almost impossible to reconcile with the short term oscillations of the field exhibited by the solar cycle. That said, there is still no consensus regarding the location or the underlying physical mechanism of the solar dynamo.

Hydromagnetic dynamos can be classified in a number of different ways; for the solar dynamo it is helpful to make the distinction between \textit{large-scale} and \textit{small-scale} dynamo action. A large-scale dynamo may be categorised as one for which the magnetic field contains a significant proportion of its energy on scales greater than that of the motions responsible for generating the field (in a turbulent flow this would be a typical eddy size); a small-scale dynamo, by contrast, has the bulk of its energy on scales comparable with, or smaller than, that of the velocity. Although such a distinction is not clear-cut, it is nonetheless instructive in providing a framework in which to discuss various types of generation mechanism. The magnetic field of the solar cycle has a spatial scale comparable with that of the Sun itself, and so would certainly seem to be the product of a large-scale dynamo. In contrast, the small magnetic elements of both signs that are observed on the quiet Sun (sometimes referred to as `salt and pepper field') could be generated \textit{in situ} by the small-scale granular and supergranular convection or, alternatively, could be the result of the shredding of the large-scale (solar cycle) field by the surface convection.

In this section we shall give a brief account of both the theoretical ideas underlying astrophysical dynamo theory and the major unresolved problems, in order that we can then discuss the present thinking on the solar dynamo. With the wide range of material covered in this review, it is simply not possible to give anything like a full set of references; we refer the reader to the review articles on the solar dynamo by \cite{Ossendrijver_03} and \cite{JTT_10}, and to the volume edited by \cite{DS_07}, which covers all the mathematical aspects of astrophysical dynamos.

\subsection{Mean field dynamos}
\label{Subsec:MFD}

Historically, astrophysical dynamo theory has focused on the problem of large-scale field generation; for the Sun this has meant seeking an explanation of the solar cycle. Mathematically, this involves seeking a full solution of equations 
(\ref{motion}) to (\ref{gaslaw}) for a highly turbulent flow covering a large range of spatial scales, a problem that, without approximations, is both analytically and computationally intractable. The most fruitful analytical approach has been via \textit{mean field electrodynamics}, an extremely elegant formulation of MHD turbulence that describes the evolution of the mean (large-scale) field in terms of transport tensors determined by the statistical properties of the small-scale velocity field \citep{SKR_66,KR_80}. The large-scale field ${\bf B}_0$ then evolves according to the mean induction equation
\begin{equation}
\frac{\partial {\bf B}_0}{\partial t} = \nabla \times \left( {\bf U}_0 \times {\bf B}_0 \right) +
\nabla \times \bfcalE 
+ \eta \nabla^2 {\bf B}_0,
\label{eq:mean_ind}
\end{equation}
where ${\bf U}_0$ is the large-scale velocity field and where the mean electromotive force (emf) $\bfcalE = \langle {\bf u} \times {\bf b} \rangle$, with $\langle \, \, \rangle$ denoting a spatial average over intermediate scales, and where $\bf u$ and $\bf b$ are the small-scale velocity and magnetic fields. Mean field electrodynamics is, at heart, a kinematic theory, in which the magnetic field is assumed not to influence the velocity field. Then, assuming that any fluctuating magnetic field arises solely from the influence of the small-scale velocity on the mean field, an assumption to which we shall return shortly, one may postulate the expansion \citep[see, for example,][]{Moffatt_78}
\begin{equation}
\calE_i = \alpha_{ij} B_{0j} + \beta_{ijk} \frac{\partial B_{0j}}{\partial x_k} + \ldots,
\label{eqn:emf}
\end{equation}
where $\alpha_{ij}$ and $\beta_{ijk}$ are pseudo-tensors, dependent on the properties of the velocity field and on the magnetic diffusivity $\eta$ (pseudo-tensors since $\bfcalE$ is a polar vector whereas ${\bf B}_0$ is an axial vector). Substitution for $\bfcalE$ from (\ref{eqn:emf}) into (\ref{eq:mean_ind}) then yields an equation for the evolution of the mean magnetic field. For illustrative purposes, it is instructive to consider the simplest case of isotropic turbulence, for which $\alpha_{ij} = \alpha \delta_{ij}$ and $\beta_{ijk} = \beta \epsilon_{ijk}$ (with $\alpha$ a pseudo-scalar and $\beta$ a scalar); equation (\ref{eq:mean_ind}) then takes the form
\begin{equation}
\frac{\partial {\bf B}_0}{\partial t} = \nabla \times \left( {\bf U}_0 \times {\bf B}_0 \right) +
\nabla \times \left( \alpha {\bf B}_0 \right) 
+ \left( \eta + \beta \right) \nabla^2 {\bf B}_0.
\label{eq:mean_ind_isotropic}
\end{equation}
The most striking difference between the unaveraged and averaged induction equations ((\ref{induction}) and (\ref{eq:mean_ind_isotropic})) is the appearance of the term involving $\alpha$ (the `$\alpha$-effect' of mean field dynamo theory). It makes explicit the possibility of dynamo action through its ability both to generate poloidal magnetic field from toroidal and, conversely, toroidal field from poloidal. Furthermore, since $\alpha$ is a pseudo-scalar, it follows that it can be non-zero only for flows that, on average, lack reflectional symmetry. These are typically characterised by having non-zero \textit{helicity} (the correlation between the flow velocity and the vorticity), a natural consequence of rotating flows. From equation~(\ref{eq:mean_ind_isotropic}), it can be seen that the $\beta_{ijk}$ term, in its simplest isotropic form, can be thought of as a turbulent diffusivity; in general though, for non-isotropic turbulence, its physical interpretation is much more complicated \citep{KR_80}.

A mean field dynamo in which both parts of the dynamo cycle are driven by $\alpha$-effects (i.e.\ in which the second term on the right hand side of equation~(\ref{eq:mean_ind_isotropic}) dominates the first term) is known as an $\alpha^2$-dynamo. The first term on the right hand side of equation~(\ref{eq:mean_ind_isotropic}) represents the advection and stretching of the large-scale field by the large-scale flow. For many astrophysical bodies, and in particular for the Sun, the generation of toroidal magnetic field is ascribed to the stretching of the poloidal field by the differential rotation of the body (known as the `$\omega$-effect'). A mean field dynamo in which the large-scale poloidal field arises from the $\alpha$-effect and the large-scale toroidal field from the $\omega$-effect is known as an $\alpha \omega$-dynamo. The prototypical $\alpha \omega$-dynamo model was first described by \cite{Parker_55}, who demonstrated how a radially dependent $\omega$-effect would lead to latitudinally propagating dynamo waves, as observed on the Sun. It should be noted that although an $\alpha^2$ dynamo can categorically be described as a large-scale dynamo (since all the motions are small-scale), such a designation is less precise for an $\alpha \omega$-dynamo, which has a combination of small-scale motions (contributing to $\alpha$) and a large-scale differential rotation $\omega$ --- although $\omega$ of itself cannot lead to dynamo action.

Unfortunately, only for the case when the magnetic Reynolds number $Rm$ is small --- which is never realised astrophysically  --- is it is possible to solve explicitly for the transport tensors $\alpha_{ij}$ and $\beta_{ijk}$. Thus the most common approach to astrophysical mean field dynamo modelling is to adopt plausible, physically motivated expressions for $\alpha_{ij}$ and $\beta_{ijk}$ (and $\omega$ if this is not known) and to explore the nature of the resulting dynamo action. Via this approach, it is also possible to model the nonlinear back-reaction of the magnetic field on the flow (via the Lorentz force) through modification of the mean field transport coefficients. Through a judicious choice of coefficients it is then possible to replicate a whole range of astrophysical magnetic fields. It is though worth sounding the note of caution that such an approach, although physically plausible, does not emanate directly from the governing equations of MHD.

\subsection{Small-scale dynamos}
\label{Subsec:SSD}

Whereas large-scale dynamo action requires organisation of the field by a large-scale emf, the key part of which is typically an $\alpha$-effect, dependent on the flow lacking reflectional symmetry, small-scale turbulent dynamo action relies on the local stretching and constructive folding of the field overcoming dissipative effects. Small-scale dynamos have mainly been studied within the framework of the \textit{fast dynamo problem}, which investigates the possibility of (kinematic) dynamo action in the limit of $Rm \to \infty$; a comprehensive account of the fast dynamo problem can be found in the monograph by \citet{CG_95}. \citet{DO_93} derived an expression for the fast dynamo growth rate, dependent on the Lyapunov exponents of the flow (reflecting the ability of the flow to stretch field exponentially) and the cancellation exponent, encapsulating the folding and dissipative effects of the flow. Given that one expects diverging fluid particle paths (Lagrangian chaos) in a turbulent flow, it is therefore natural to associate small-scale dynamo action with any suitably turbulent flow. This is borne out by a number of numerical simulations of MHD turbulence \citep[e.g.][]{Schekochihin_etal_07}. One issue though that is not fully resolved is whether small-scale turbulent dynamo action can persist at very small magnetic Prandtl number, $Pm$. This is a difficult problem to address numerically; the indications are however that dynamo action does survive at small values of $Pm$ \citep{BC_04}.

\subsection{How do the two dynamo mechanisms interact?}
\label{Subsec:interact}

The theoretical treatment of large- and (fast) small-scale dynamos has developed along independent lines, reflecting the different physical processes needed for each type of dynamo action; a lack of reflectional symmetry in the flow for the former, a suitably chaotic flow for the latter. However, whereas such a distinction is convenient for dynamo theorists, in reality, a turbulent, rotationally-influenced astrophysical flow will possess the necessary ingredients for both large- \textit{and} small-scale dynamo action. Given this, it is important to look again at the mean field formulation so as to understand the impact of small-scale dynamo action. As mentioned above, the crucial point is that the expansion (\ref{eqn:emf}) assumes that any fluctuating field results entirely from distortion of the large-scale field; this is equivalent to saying that the small-scale field would decay of its own accord --- i.e.\ no small-scale dynamo. However, if there \textit{is} a small-scale dynamo then, at least kinematically, the fluctuating field will grow according to the local stretching and folding properties of the flow. \cite{BCR_05} and \cite{CH_09} have argued that, under these circumstances, the growth rate for the small-scale dynamo will dominate that of a mean field dynamo (certainly in the case of an $\alpha^2$-dynamo) and that consequently the resulting field will result from small-scale dynamo action, with any large-scale component simply a result of the `spilling out' of the small-scale eigenfunction to larger scales, and \textit{not} governed by an equation such as (\ref{eq:mean_ind_isotropic}). In other words, the idea of building up a large-scale field from a weak seed field, via the standard ideas of mean field electrodynamics, seems problematic. That said, large-scale astrophysical magnetic fields do, of course, exist, and it is therefore important to understand how these may be generated, and how the problems of mean field theory may be circumvented.

One of the great unknowns in large-scale field generation is the precise role of large-scale velocity shear. \cite{HP_09} have shown how the addition of a large-scale shear to a rotating convective flow that has a weak $\alpha$-effect but no dynamo (large-scale or small-scale) can indeed facilitate large-scale dynamo action. There are various mechanisms by which this might occur. Even within the mean field framework there are three possibilities: one is simply that a vigorous $\omega$ can compensate for a feeble $\alpha$; another is that the existing $\alpha$-effect is `improved' by the introduction of large-scale coherence; and a third is that there can be a more complicated mean field mechanism, essentially through modification of the non-diffusive elements of the $\beta_{ijk}$ tensor \cite[e.g.][]{RK_03,Proctor_07}. A very different possibility is that the large-scale shear, although not of itself capable of dynamo action, interacts with the small-scale convective flow to produce a flow that, entirely through its large scales, can now act as a dynamo; such a dynamo would then be a `small-scale' dynamo (field and flow on the same scales) but at the large scales! The important task of untangling the various possible mechanisms is, however, far from straightforward; the recent efforts of \cite{PH_11} proved inconclusive. It is also far from clear whether the addition of a large-scale shear flow to flows that act as healthy small-scale dynamos (which is probably a truer reflection of astrophysical flows) will generate a significant large-scale magnetic field. Fully understanding the interplay between large- and small-scale flows in a highly turbulent flow remains one of the great challenges of the solar dynamo problem.

\subsection{The dynamo(s) in the Sun}
\label{Subsec:solar_dynamos}

The first point to address is whether two separate dynamo mechanisms are operating in the Sun; i.e., is the small-scale field observed at the solar surface generated locally by the granular and supergranular convection, or is it simply due to the shredding of the large-scale, solar cycle field? On theoretical grounds, we would certainly expect local small-scale dynamo action, since flows in the surface regions are turbulent and have a high value of $Rm$. Observationally, there are two factors that point to the existence of a small-scale dynamo. One is that the small-scale surface field is not correlated with the solar cycle, as discussed earlier; the other is that observations suggest a renewal time for magnetic flux in the quiet Sun on a timescale of less than a day \citep{Hagenaar_etal_2003}. So, most probably, there are indeed two distinct dynamo mechanisms operating in the Sun.

Notwithstanding the possible problems discussed in Section~\ref{Subsec:MFD}, the large-scale solar dynamo is generally discussed within the framework of mean field $\alpha\omega$-dynamos; three very different models have been quite widely discussed --- though it should be stressed that there is no guarantee that the solar dynamo necessarily falls into any of these possibilities. Here we shall discuss the pros and cons of these various models before looking into an alternative, less widely studied, possibility.

In some sense, the most natural explanation of the solar dynamo, certainly without any detailed knowledge of the internal flow dynamics, is that it is a `distributed dynamo', with $\alpha$ and $\omega$ distributed smoothly throughout the convection zone; such a dynamo could occur wherever rotation and convection take place together. There are, however, some major difficulties with this idea. One is that concentrated magnetic field is buoyant and hence will escape from the convection zone on a timescale much shorter than the $11$ year generation time of the dynamo \citep[see][]{Parker_79}. Furthermore, \cite{GW_81} pointed out that weaker fields also would be expelled, in this case by convective motions. Thus the idea was advanced of a dynamo located at or just beneath the convective zone \citep{SW_80, GRVW_81}. Several years later, further support for the idea of such a deep-seated dynamo came from the helioseismological deductions of the internal rotation; the strong radial gradient of differential rotation in the tachocline makes this an obvious location for the $\omega$-effect.

The most popular current explanation of such a deep-seated dynamo is as an \textit{interface dynamo}, in which the two elements of the dynamo cycle (`$\alpha$' and `$\omega$' in simple terms) are spatially separated, an idea first proposed by \cite{Parker_93}. The $\alpha$-effect is envisaged to result from helical motions at the base of the convection zone, whereas the $\omega$-effect comes from the differential rotation in the tachocline. Such a dynamo makes natural use of the radial variation of $\omega$ in the tachocline, though suffers from no proper knowledge of how a significant $\alpha$-effect may be generated, nor how the two parts of the dynamo cycle may be realistically coupled.

For both of the two dynamo mechanisms described above, sunspots and active regions are `by-products' of the dynamo process (`epiphenomena' in the terminology of \cite{Cowling_75}) --- albeit ones of great interest --- but are not critical to the operation of the dynamo. By contrast, in the third model of the solar dynamo that is currently widely discussed, the surface field is inherent to the dynamo process. As in the interface dynamo, the \textit{flux transport} dynamo has spatially separated $\alpha$ and $\omega$ effects, but now they are assumed to be widely separated, by the entire depth of the convection zone. The $\omega$-effect is again supplied by the differential rotation in the tachocline, with the $\alpha$-effect resulting from the dispersal of the surface poloidal field acting to cancel out the field of the previous cycle and initiate a new poloidal field with a change of sign --- this idea dates back to the ideas of \cite{Babcock_61} and \cite{Leighton_69}. In its modern incarnation, the two components of the dynamo cycle are linked by a meridional flow, which acts to transport poloidal field from the solar surface to the tachocline, where it is sheared out to produce toroidal field \citep[e.g.][]{CSD_95}. The flux transport dynamo has the pleasing feature, at least to external observers, that part of the dynamo cycle can actually be observed. It does though rely crucially on a meridional circulation that  (i) extends across the entire depth of the convection zone, and (ii) is able to transport a magnetic field unscathed through the highly turbulent convection zone. Neither of these are particularly plausible. Indeed, very recent observations by \cite{Hathaway_2011} of the motions of supergranules suggest that the poleward meridional flow observed at the surface has a very shallow return (equatorward) flow, at a depth of only 35\,Mm. If this is indeed the case, then it deals a fatal blow to the idea of a flux transport dynamo.

Thus, all three current models have serious difficulties, and indeed the answer might lie elsewhere. One possibility is a variation on the interface dynamo discussed above, in which the electromotive force needed to generate poloidal field comes not from convection but from a magnetic buoyancy instability of the magnetic field \citep{Schmitt_84, Thelen_2000a, Thelen_2000b, DH_11}. The instability would then play two roles: to release magnetic flux from the tachocline, eventually to appear as active regions, and also to generate an electromotive force, which could close the dynamo cycle. A dynamo of this form is inherently nonlinear, since the electromotive force arises only from a pre-existing field; there is no kinematic, or weak field, regime. Such a dynamo mechanism escapes some of the problems of obtaining an $\alpha$-effect from turbulent convection; although no fully convincing such model has yet been developed, the idea certainly deserves further study.

\section{Comparison Between the Dynamos of the Sun and the Earth}\label{sec:comparison}

The aim of this section is certainly not to provide a review of the geodynamo problem (see, for example, the comprehensive reviews by \cite{RG_2000, Glatzmaier_2002, Fearn_2007}), but rather to discuss the similarities and differences between the geodynamo and solar dynamo; both in terms of the dynamos themselves and also in the attempts to understand them theoretically and computationally. It may be regarded as being complementary to the more detailed comparison between terrestrial and solar dynamos conducted by \cite{ZS_2006}.

One of the most striking differences between the magnetic fields of the Earth and the Sun concerns their variability in time. Whereas the geomagnetic field is characterised by long periods of one polarity, interspersed with fairly rapid periods of field reversal, the solar field changes sign on a timescale of only about $11$ years (though with modulation of the whole cycle, as discussed in Section~\ref{sec:sc}). For both bodies, but for different reasons, it is difficult to infer through surface observations the structure of the magnetic field in the region in which it is generated. For the Earth, it is possible, at least in theory, to deduce the potential field in the mantle (assuming it is a perfect insulator) through knowledge of the surface field. Through the continuity of the normal field, this would then convey information about the radial field at the top of the core. However, modes with wavenumber greater than about $15$ are dominated by the crustal magnetic field, and so it is only possible to say anything about low order modes. For the Sun, it is now possible to resolve surface and coronal magnetic features on scales down to a few hundred kilometres; the difficulty, in terms of the dynamo, is trying to relate this activity to the seat of dynamo activity, which is probably located beneath the convection zone.
  
In terms of the energy source for the dynamos, the picture is much clearer for the Sun. Whatever the precise dynamo mechanism, the energy source is rotationally-influenced thermal convection (even in the flux transport dynamo, for which the convection seems to play no obvious role, the meridional circulation is a consequence of convection in a rotating frame). In the Earth, possible energy sources are thermal convection, compositional convection or precession. The consensus seems to be that the most likely energy source is compositional convection, resulting from solidification at the inner core and the subsequent release of light elements. However, this is a complex and imperfectly understood process, and in most dynamo models standard thermal convection is employed. In the Sun, the difficulty is not in identifying the source of energy, but rather in ascertaining the nature of turbulent, compressible, rotating convection at the parameter regimes that pertain in the Sun. For both bodies it is certainly fair to say that the nature of the motion that drives the dynamo is far from clear.

Quantitative differences between the geodynamo and the solar dynamo are provided by comparisons between the dimensionless parameters describing each. The geodynamo is characterised by being turbulent ($Re = O(10^9)$) and strongly constrained by rotation (Ekman number $E = O(10^{-15})$); it does though operate at a rather modest magnetic Reynolds number ($Rm = O(10^2$)).  The solar dynamo, on the other hand, is less rotationally constrained, but operates at very high values of both Reynolds numbers ($Re = O(10^{13})$, $Rm = O(10^{10})$).
Even with the most powerful computers, it is impossible to perform dynamo simulations at parameter values even remotely close to their realistic values. Thus, in models of the geodynamo, $E$ is orders of magnitude too large, with $Re$ similarly too small, whereas in solar dynamo simulations the problem is that both the Reynolds numbers are too small; the usual approach is to push these as high as possible, currently $O(10^5)$, though this means operating at $Pm$ of order unity, which is also unrealistic. Thus, the unavoidable compromises result in computational models of the geodynamo and solar dynamo being much more similar than the dynamos themselves.

Although, historically, both the geodynamo and solar dynamo have been explored via mean field dynamo models, most of the current thrust of research is through numerical simulations, employing either spherical models, designed to capture the global magnetic field, or more local plane layer models, aimed at understanding specific physical process in detail. In terms of global modelling, it seems that geodynamo models are closer to reality than their solar counterparts. The most ambitious models are those of Glatzmaier and Roberts (1995, 1996; and other papers reviewed in Roberts and Glatzmaier, 2000; Glatzmaier, 2002), which describe dynamo action in a spherical shell driven by rapidly rotating convection. The early models are Boussinesq and consider only thermal buoyancy; the later models employ the anelastic approximation and also consider compositional buoyancy. The success of the simulations is that the resulting field is Earth-like in that it undergoes polarity excursions and dipole reversals. Of necessity in such a computational undertaking, as discussed earlier, approximations have to be made. \cite{GR_95, GR_96} adopt what is known as \textit{hyperdiffusion}, which, for their models, weights the various diffusivities according to spatial scales, leading, for example, to a smaller Ekman number at the larger scales than the smaller scales. This allows exploration of the small Ekman number regime, though at the possible risk of distorting the dynamics of the dynamo, as discussed by \cite{ZJ_97}.

Global models of the solar dynamo date back to the pioneering work of \cite{Gilman_Miller_81}, who considered a rotating, convective, Boussinesq fluid, and \cite{Glatzmaier_84, Glatzmaier_85}, who employed the anelastic approximation. These models demonstrated, for the first time, the possibility of dynamo action without the need for any parameterisation such as an $\alpha$-effect. The most recent global modelling of the solar dynamo has been performed with the ASH (Anelastic Spherical Harmonic) code \citep{Clune_etal_99}, which evolved from the  earlier anelastic code of Glatzmaier. Thus, interestingly, and reinforcing the earlier comment that computational models of the geo and solar dynamos are not as dissimilar as the dynamos themselves, two of the most widely used codes for modelling these two dynamos have a common ancestry. The ASH code results of \cite{Brun_etal_2004} again demonstrated the viability of self-consistent dynamo action by rotating convection, but showed that the resulting magnetic field was essentially small-scale, in contrast to that of the Sun. The dominance of the small-scale magnetic field was demonstrated also, in a very different model, by \cite{Livermore_etal_2007}, who considered the generation of magnetic fields by forced flows, of differing helicities, in a spherical shell.

The greater success of geodynamo models may reflect the fact that in the Earth, dynamo action is indeed distributed across the outer core, and hence is well represented by a global model, whereas in the Sun, localised generation at the base of the convection zone and in the tachocline is the crucial element. Current global models are unable to resolve all the details of the thin tachocline region; achieving a greater understanding of the solar dynamo will certainly therefore require a better understanding of the more localised physics in this region.

\section{Global Coronal Magnetic Field}\label{sec:coronalB}
The structure of the global coronal magnetic field, as it varies during the solar cycle, is important in determining where magnetic energy is stored and where it may be released in the form of solar flares, prominence eruptions and CMEs. At large heights in the corona, the magnetic field becomes sufficiently weak that the solar wind can pull the field open into interplanetary space. At this point, the field is almost radial, with the magnetic field emanating from  the north pole
separated from that of the south pole by an equatorial current sheet. The amount of open magnetic flux varies over the solar cycle, with the maximum occurring around two years after the maximum number of sunspots. The open flux influences the number of cosmic rays hitting Earth's atmosphere and so the variation in the amount of open flux during the solar cycle is linked to the variation in cosmic rays reaching the Earth. 

Modelling the evolution of the global coronal magnetic field requires knowledge of the evolution of the photospheric magnetic field
in order to provide the correct boundary condition on the photospheric surface. Unfortunately, this field is not easy to impose. Ideally, one 
would use the observed magnetic field from magnetograms but the resolution of the global numerical calculations is much 
coarser than the observations and some averaging is needed. As far as the global coronal field is concerned, it is the active region 
magnetic fields that dominate the large-scale structure. Ideally, a numerical simulation should combine the evolving
photospheric magnetic field with the evolution of the global coronal field. A model that provides the photospheric radial component of 
the magnetic field as a suitable boundary condition is now described.

\subsection{Flux transport model of the photospheric magnetic field}\label{sec:FTmodel}
\begin{figure}
\centering
\includegraphics[width=0.9\textwidth]{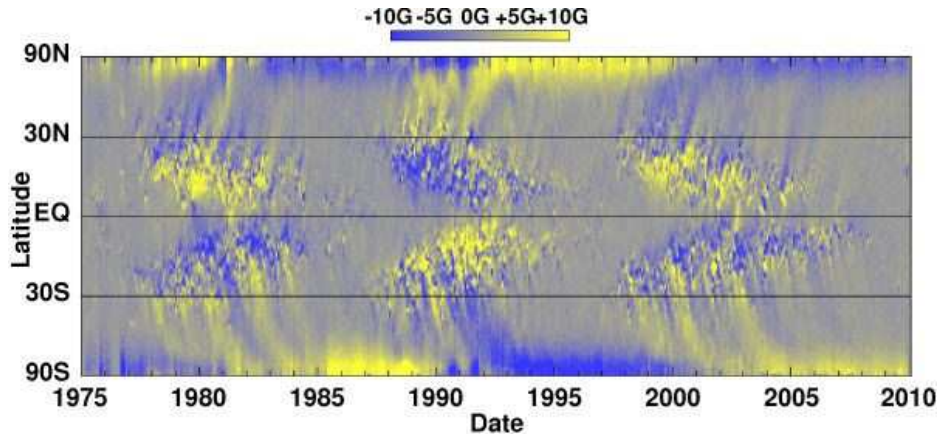}
\caption{The solar butterfly diagram \citep[from][]{Hathaway2010}, where yellow represents positive magnetic flux and blue negative flux, on a scale between $\pm 10$G.}
\label{hathaway}
\end{figure}

The nature of the solar cycle becomes clear when the line-of-sight magnetic field, averaged in longitude, is plotted as a function of latitude and time (see Figure~\ref{hathaway}). What is clear is that the evolution of the magnetic field is different in the low latitudes, between $\pm 30^\circ$, compared with the high latitudes, above $\pm 30^\circ$. In the low latitudes, the sunspots of a new solar cycle emerge at around $\pm 30^\circ$. Here regions of strong magnetic field may last for around two weeks before dispersing. As the cycle advances, new sunspots emerge closer and closer to the equator. In contrast, at higher latitudes, there is a slow drift of the trailing polarity magnetic flux towards the polar region. This is \textit{not} due to the emergence of sunspots but instead to a slow, observed plasma flow called the \textit{meridional flow}. This drifting field is of opposite polarity to the field in the polar regions. Eventually it cancels with the polar field, creating a reversal in the sign of the polar polarity. Note that the reversal in the polar polarity occurs around one to two years after cycle maximum. How or if this is linked to the solar dynamo is unknown (see Section~\ref{Subsec:solar_dynamos}).

Once a new sunspot pair has emerged, it evolves in response to three main effects. First, the two sunspots tend to appear at slightly different latitudes, with the leading spot being closer to the equator than the following spot. The two sunspots appear to drift apart owing to the photospheric differential rotation, discussed above. Second, there is the slow meridional flow from the equator to the poles, with, presumably, a return flow somewhere in the solar interior \citep[see][]{Hathaway_2011}. Third, towards the end of the sunspots' lifetime, they begin to spread out, with supergranular flows producing a random walk of the dispersed magnetic elements. This random walk can be described by a diffusion process. Each of these three effects is now discussed.

\subsection{Large-scale surface flows}
Observations of the rotation speed of photospheric features indicate that the surface rotation of the Sun can be described by \citep[see][]{Snodgrass1983},
\begin{equation}
\Omega (\theta) = 13.38 - 2.30 \cos^2 \theta - 1.62 \cos^4 \theta \hbox{  deg day}^{-1},
\label{diffrotation}
\end{equation}
where $\theta$ is the co-latitude. The meridional flow profile may be specified as
\begin{equation}
u(\theta) = \left \{\begin{array}{cc}  -u_0 \sin (\pi \lambda/\lambda_0), & |\lambda| < \lambda_0,\\
            0, & \hbox{otherwise},
          \end{array}\right .
\label{meridional}
\end{equation}
where $\lambda = \pi/2 - \theta$ is the latitude and $\lambda_0 = 75^\circ$ is the latitude above which the meridional flow vanishes. The typical flow speed, $u_0 = 11$m s$^{-1}$, is taken from Hathaway (1996).

The diffusion of active region magnetic fields can be described by a uniform diffusivity coefficient
$D = 200 - 450$\,km$^2$\,s$^{-1}$. The diffusion timescale, based on the value of $D$ and
global lengthscales, is of the order of 34 years. This makes diffusion the slowest of the physical processes involved in the evolution of the photospheric magnetic field over the solar cycle. In contrast, differential rotation is the fastest process with a timescale of around a quarter of a year. 

\subsection{A flux transport model}
Using the three physical processes above, fitted to observed data, it is possible to model the evolution of  the radial component of the photospheric magnetic field over the solar cycle \citep[see][]{Wang1989, Sheeley2005, Mackay2006}. The key ingredient is insertion of new active regions, taken from their observed properties of flux, latitude and tilt. The long time evolution of $B_r$ at the solar surface is based on the radial component of the induction equation~(\ref{induction}),
\begin{equation}
\frac{\partial B_r}{\partial t} = \frac{1}{\sin \theta}\frac{\partial}{\partial \theta}\left [\sin \theta \left (-u(\theta) B_r + 
D\frac{\partial B_r}{\partial \theta}\right )\right ] - \Omega(\theta)\frac{\partial B_r}{\partial \phi} + \frac{D}{\sin^2 \theta}\frac{\partial^2B_r}{\partial \phi^2}
\label{FTM.eq}
\end{equation}
\citep[see][for details]{Wang1989}.
Thus, the surface distribution of the radial component of the coronal magnetic field can be determined and used as input to models that predict the global structure of the coronal field. While this model governing the transport of the surface magnetic flux has a certain simplicity to it, it has been successful in reproducing (i) the strength of mid-latitude fields, (ii) the reversal of the polar fields and the timing of the reversal after the start of the new cycle, (iii) the origin of certain repeating 28-29 day patterns.

\subsection{Force-free coronal fields}\label{sec:globalB}
Since the plasma $\beta$ (the ratio of gas to magnetic pressure) in the low solar corona is small, it is normal to model the global coronal magnetic field as a \textit{force-free} field. It can be modelled by a variety of different schemes of varying complexity. The simplest is the potential field model, in which the field is described by a magnetic potential satisfying Laplace's equation \citep{Schatten1969}. Next, more generally, is the force-free field model where ${\bf j} \times {\bf B} = {\bf 0}$. This implies
\begin{displaymath}
\nabla \times {\bf B} = \alpha {\bf B}, \qquad \hbox{and} \qquad {\bf B}\cdot \nabla \alpha =0;
\end{displaymath}
$\alpha = 0$ gives the potential field, $\alpha = \hbox{constant}$ corresponds to the linear force-free field. In general, $\alpha$ is constant along a magnetic field line, but if it varies from field line to field line then the result is a non-linear force-free field \citep{Schrijver2003}. [Note that in this section, following standard notation, $\alpha$ is related to the component of the current parallel to the magnetic field and $\beta$ is the ratio of gas to magnetic pressures. In Section~\ref{Subsec:MFD} (also in standard notation!), $\alpha$ and $\beta$ were pseudo-tensors in the expansion of the mean electromotive force.]

Boundary conditions must be specified before the magnetic field can be determined. At the solar surface ($r = R_\odot$) the observed radial magnetic field values are used and at an outer radius 
(normally taken as $2.5 R_\odot$) the magnetic field is assumed to be purely radial. This latter condition simulates the effect of the solar wind opening the field. These models are called \textit{force-free source surface} models.

Vector magnetograms \citep{Metcalf1995} provide information about the horizontal magnetic field components and estimates of the spatial variation in $\alpha$. What is clear is that the photospheric field has non-zero $\alpha$ and, thus, the coronal field  cannot be represented by simple potential models. Modelling non-linear force-free fields is much more complicated, and models that use magnetic field data from a single magnetogram are very limited. To make real progress, it is essential to track the movement of the magnetic footpoints at the photosphere in response to the differential rotation, the meridional flow and the diffusion that describes the breakup of active regions. Non-linear force-free fields, based on the photospheric magnetic field distribution described by the surface flux transport model are particularly powerful at predicting
various coronal phenomena \citep{Mackay2006}.

\subsection{Results from global field models}
The global coronal magnetic field model of \cite{Mackay2006} has successfully predicted some important features that can be tested by observations. In agreement with observations, it has been shown that (i) the open magnetic flux reaches a maximum about two years after solar maximum, and (ii) it is the non-potential nature of the coronal magnetic field that is responsible for the inflation of the corona and the opening of the closed magnetic field \citep{Yeates2010}. 

The global model has also successfully predicted many properties of \textit{solar filaments} or, synonymously, \textit{solar prominences}. Prominences consist of cool dense plasma, supported high in the hot tenuous corona by the local magnetic field. They are very narrow (width around 5\,Mm) and extremely long structures (lengths up to 700\,Mm). They always lie above the polarity inversion line, a line observed in magnetograms that separates photospheric regions of positive and negative polarity. Solar prominences tend to form when the local magnetic field has a dip (a location where the magnetic tension force acts vertically upward). In addition, the direction of the dominant component of the magnetic field along the prominence is called the \textit{chirality}. Prominences in the northern hemisphere are predominantly \textit{dextral} while those in the southern hemisphere are predominantly \textit{sinistral} \citep{Mackay2010}. However, there are a few exceptions to this hemispheric pattern. The global model is able to predict the location of prominences and their chirality with up to 96\% accuracy, including
the hemispheric exceptions \citep{Yeates2008}. In addition, the loss of a stable equilibrium, which the code can detect, can be linked to the initiation of a CME. The location of a CME can be predicted with 50\% accuracy \citep{Yeates2009}, something that no other simulation code is able to do.

\section{Flux Emergence}\label{sec:emergence}
\subsection{Numerical Simulation Results}
Recently, numerical simulations have been used to investigate how the magnetic field emerges from the solar interior. An initial magnetic structure is chosen and its subsequent evolution
tracked. Presently, the exact form of the initial interior magnetic field is unknown, although
helioseismology does indicate the presence of sub-photospheric fields. In principle, this should be given by the results of dynamo simulations but at present there is no such input. In the absence of accurate observational data, the initial magnetic field is chosen as either a twisted cylindrical flux tube \citep{Fan2001, Archontis2004} or a twisted toroidal flux tube \citep{Hood2009,MacTaggart2009}. This seems reasonable since sunspots appear to consist of a large flux tube. In addition, \cite{Emonet1998} and \cite{HFJ_98} have shown that a magnetic flux tube rising in the solar convection zone must be twisted in order that it is not pulled apart by the turbulent convective flows.

\begin{figure}
\centering
\includegraphics[width=0.45\textwidth]{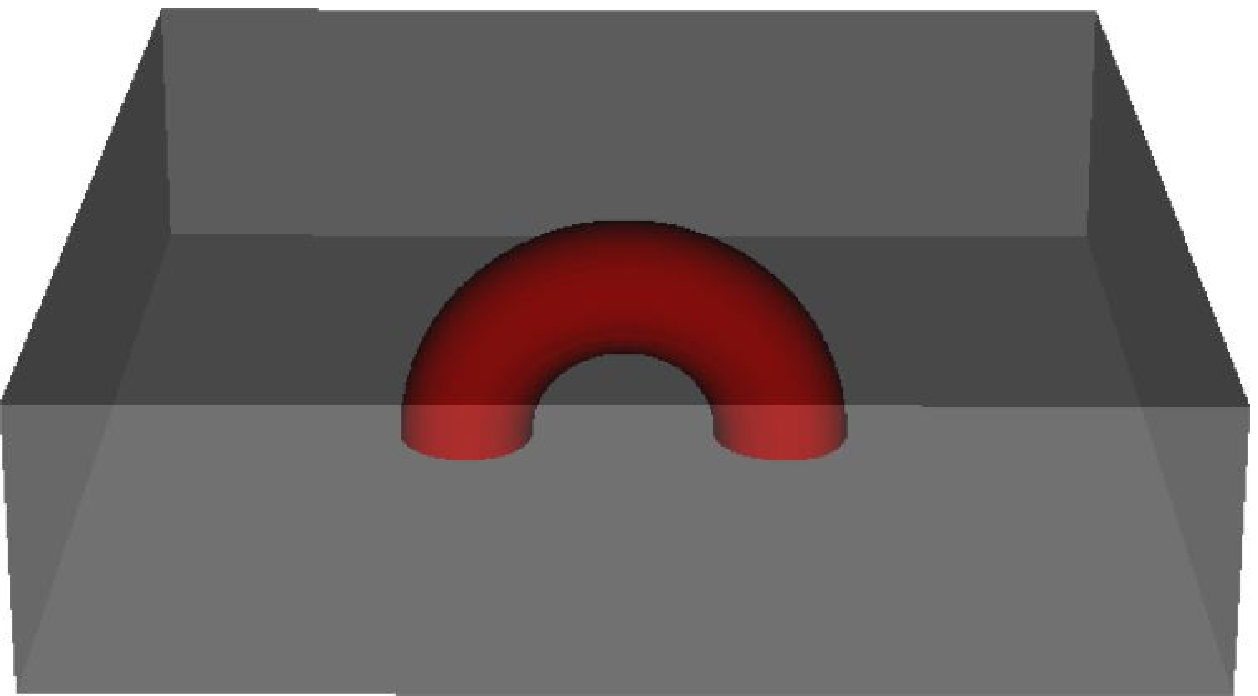}
\includegraphics[width=0.45\textwidth]{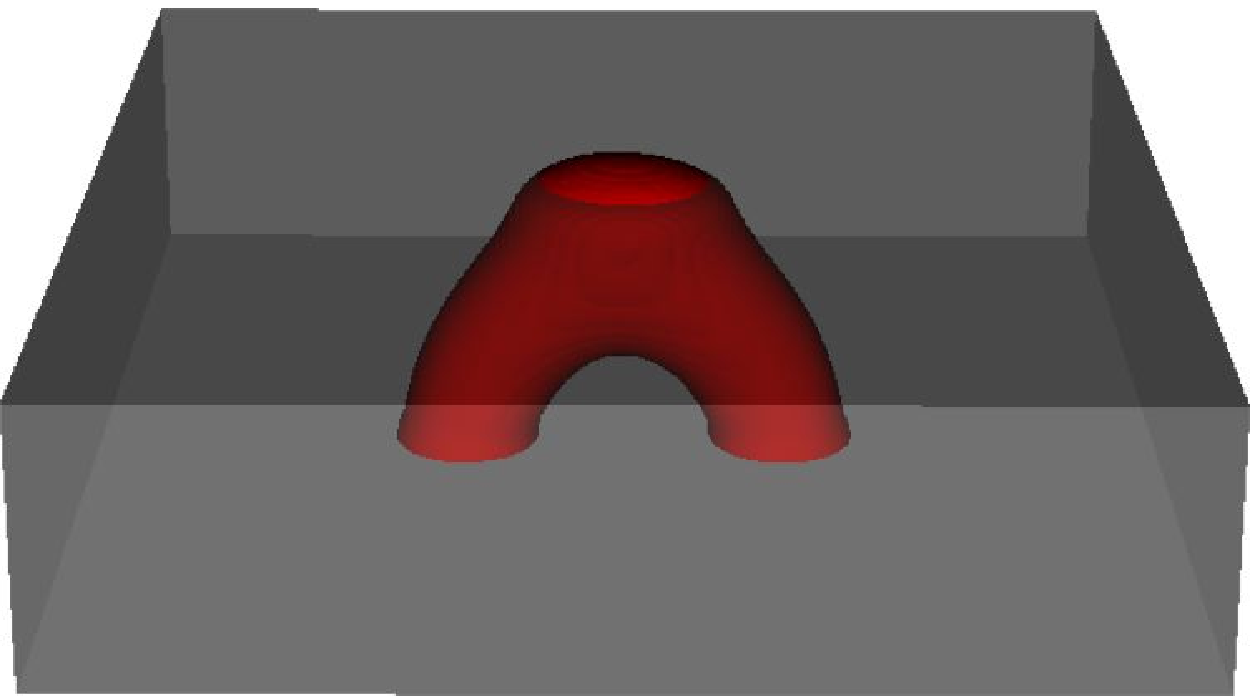}
\includegraphics[width=0.45\textwidth]{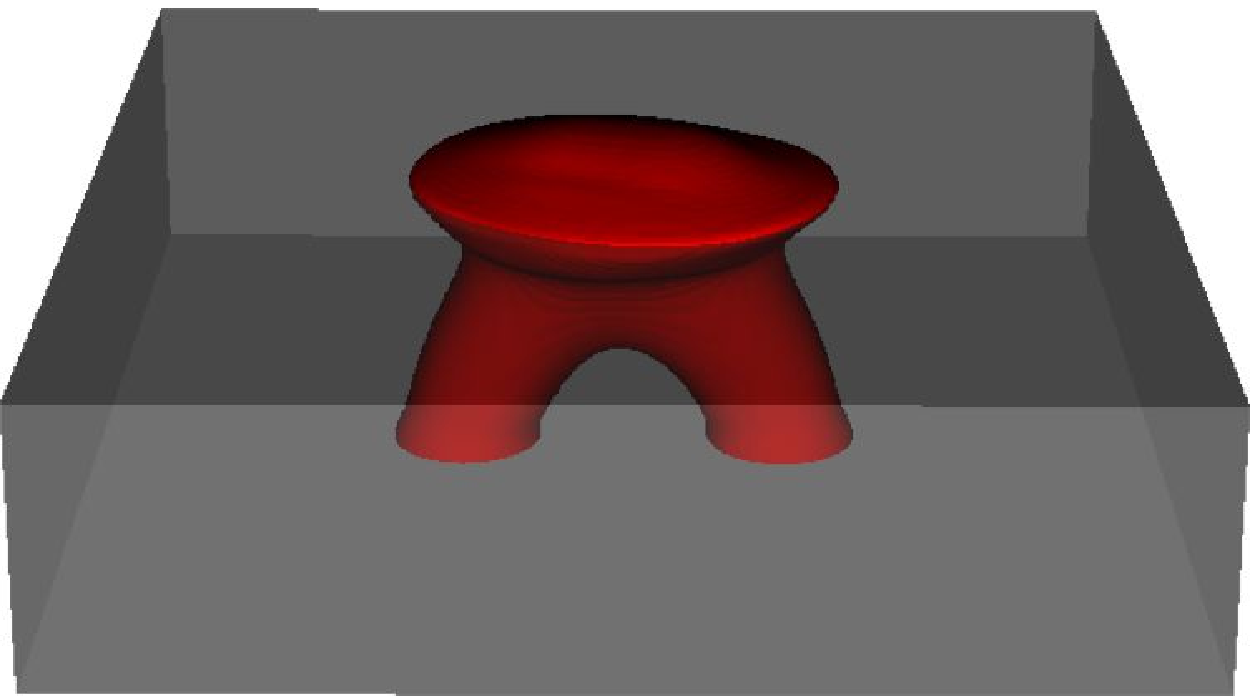}
\includegraphics[width=0.45\textwidth]{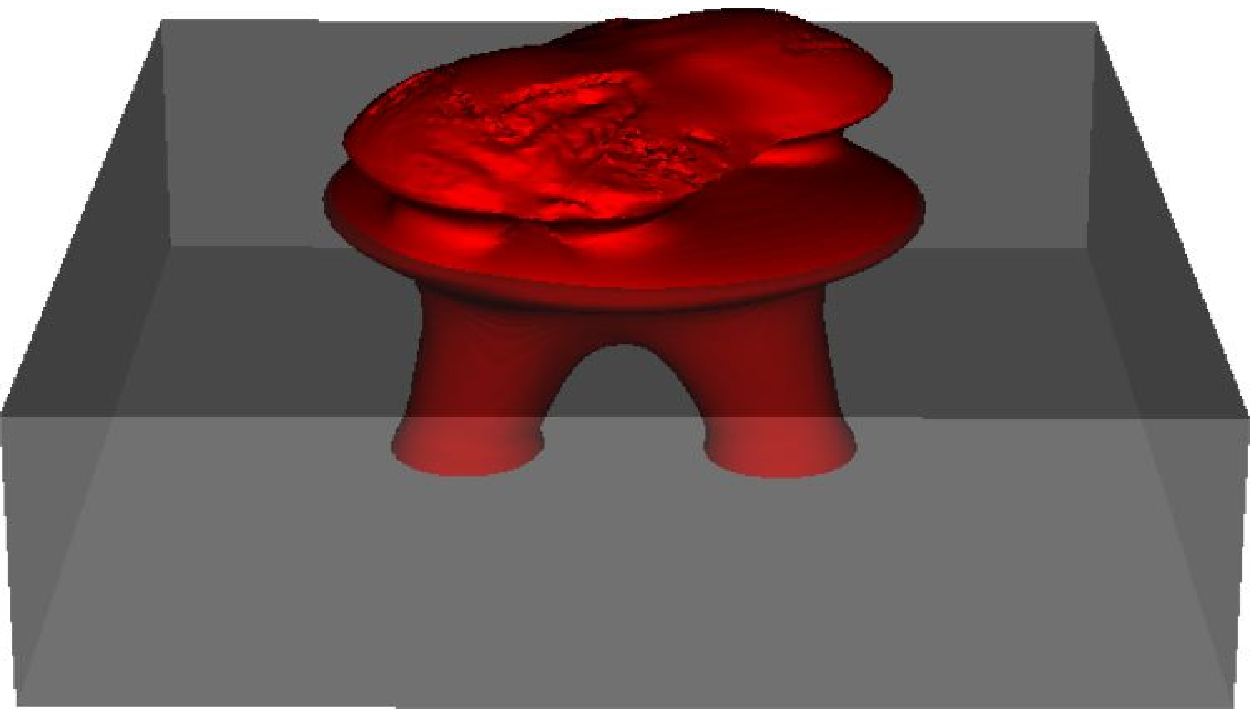}
\caption{The isosurface of the magnetic field magnitude shows the initial state (top left). The grey region represents the solar interior. The initial tube is buoyant and rises to the photosphere (top right). Since the atmosphere is stably stratified, the rise stops and the field spreads out (bottom left) until the magnetic buoyancy instability is triggered (bottom right).}
\label{fig:magbuoyancy}
\end{figure}

Consider a flux tube near the top of the solar convection zone. If the tube is in total horizontal pressure balance (i.e.\ gas pressure plus magnetic pressure), then the gas pressure inside the tube must be less than the surrounding external gas pressure. Now if the tube is in horizontal temperature balance, then using the ideal gas law, the density inside the tube must be less than the density in the exterior. The tube is light and begins to rise owing to buoyancy. Since the solar convection zone is unstably stratified, this rise will continue unchecked. The plasma $\beta$ is large in the interior and so the magnetic flux tube is simply carried along by the rising plasma. This is illustrated in Figure~\ref{fig:magbuoyancy} (top right). 

However, once the magnetic field reaches the photosphere, the atmosphere is stably stratified. Thus, the magnetic field is trapped at the solar surface and starts to expand horizontally as shown in Figure~\ref{fig:magbuoyancy} (bottom left). However, the field in the interior continues to push towards the photosphere, increasing the magnetic field strength at the solar surface until a magnetic buoyancy instability is triggered. Only now is it possible for the magnetic field finally to rise into the corona (Figure~\ref{fig:magbuoyancy}, bottom right). The condition for the onset of the magnetic buoyancy instability is presented in \cite{Acheson1979} and the application to emergence at the solar photosphere is discussed in \cite{Archontis2004}.

Two points are worth noting. First, there is a large amount of magnetic flux that remains trapped in the interior, even after the emergence process has occurred. Second, the emergence due to the magnetic buoyancy instability starts once the gradient in the magnetic field has increased and the plasma $\beta$ has dropped to order unity. Hence, the Lorentz force is now comparable with the pressure gradient force and the magnetic forces are strong enough to move the plasma. This is different to the situation in the solar interior and is important when interpreting the consequences of emerging fields, as discussed below.
\subsubsection{Sunspots}
The first magnetic field to emerge in numerical simulations always appears in the North-South direction, but quickly spreads apart into the correct East-West direction. This is due to the twist in the magnetic field, needed to keep it coherent in the solar interior. As the flux tube starts to rise in the interior, it expands and the azimuthal magnetic field component is enhanced as compared with the axial component. Thus, the outer sections of the flux tube become more twisted and the first fields to emerge are perpendicular to the main axis of the tube.

The magnetic field, when it expands into the corona, becomes simpler in structure and rapidly relaxes towards a force-free state. The field remaining in the interior is still twisted and torsional Alfv\'en waves propagate from the interior up towards the untwisted coronal field. The net effect at the photosphere is a rotation of the sunspots and, indeed, rotating sunspots are frequently observed.

The sunspots in active regions tend to drift apart but eventually reach a final separation, before they start to break up and disperse. This final separation distance of sunspots depends on how the tube becomes buoyant in the interior and where in the interior the ends of the flux tube are rooted. The toroidal model has a fixed separation distance given by the major diameter \citep[see][]{Hood2009}. This is not the case in the cylindrical models, where there must be a restriction on the size of the buoyant section \citep[see][]{MacTaggart2009a} to avoid a continual
separation of the spots.

\subsubsection{Flux rope formation}\label{subsec:rope}
The emergence process readily results in the formation of new flux ropes that then begin to erupt outwards towards the corona and may represent the initiation of a CME. This is discussed in \cite{Archontis2008}, but is seen in many other simulations \citep[for example,][]{Manchester2004}. As discussed above, the magnetic field begins to emerge through the photosphere only once the conditions for a magnetic buoyancy instability are satisfied. This requires that the magnetic pressure is comparable with the gas pressure. Since the gas pressure has a very small gravitational scale height, the background gas pressure drops off rapidly. So once the magnetic pressure exceeds the gas pressure at the photosphere, the magnetic forces cause a rapid rise of the magnetic field, removing the excess pressure and, indeed, producing a total pressure deficit. Thus, after the rapid rise of the field, there is an inflow or converging motion in the mid to high photosphere.

\begin{figure}[t]
\centering
\includegraphics[scale=0.2]{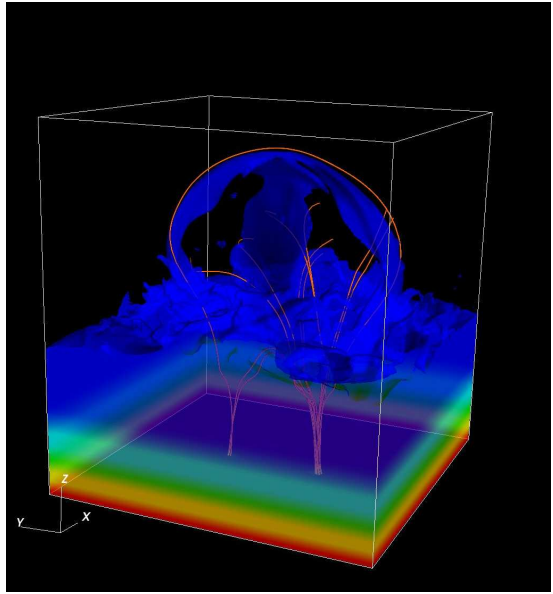}
\caption{Contours of plasma density are shown, with a few sample magnetic field lines in orange. The structure seen is erupting out towards the top of the computation box and consists of dense plasma surrounded by the magnetic field of a newly formed flux rope. (MacTaggart, private communication.)}
\label{fig:fluxropew}
\end{figure}
In addition, the rapid expansion of the emerging magnetic field, into a dipole structure, creates magnetic field lines that almost bend back on themselves. Hence, there are strong currents
at the photosphere and the resulting Lorentz force gives rise to a shearing motion on either side of the polarity inversion line.  The shearing, together with the inflow, results in inclined magnetic field lines being brought together. A current sheet forms and magnetic reconnection results in the formation of a flux rope that starts to erupt into the corona (see Figure~\ref{fig:fluxropew}). 
This new rope may be either above the original axis of the original flux tube or below it.

\subsubsection{Sigmoids}
Sigmoids are observed in X-rays in the solar corona. They take the form of forward or reverse S-shaped structures and are an indication that  the magnetic field is highly non-potential. Sigmoids have been recorded by several solar missions (e.g.\ Skylab, Yohkoh, SoHO, Hinode); the forward S-shape structures are mainly formed in the southern hemisphere, while the inverse S-shape structures are observed in the northern hemisphere \citep[see][]{Pevtsov1997}. These brightenings were named \textit{sigmoids} by \cite{Rust1996}, who also showed that 
many of the sigmoidal brightenings evolve into arcades, which are often associated with the eruption of CMEs. In general, the appearance of sigmoids in active regions is closely related to intense solar activity. Observational studies of sigmoids can be found in \cite{Canfield1999, Canfield2007}.

\begin{figure}[th]
\centering
\includegraphics[scale=0.5]{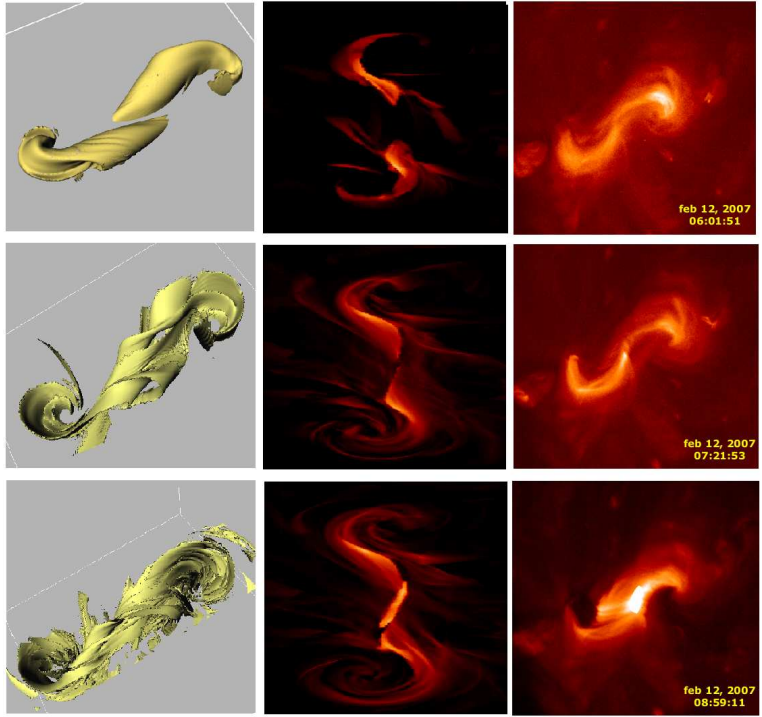}
\caption{ Left column: Isosurface of current density at three times during the simulation. Middle column: Estimate of the X-ray plasma emission from the simulations. Right column: Soft X-ray observations from XRT on Hinode. \citep[From][]{Archontisetal2009}}
\label{fig:sigmoids}
\end{figure}

Sigmoid structures naturally appear when a twisted flux rope emerges owing to the emergence of non-potential magnetic fields and the injection of magnetic helicity. The simulation results of 
\cite{Archontisetal2009} show good agreement between the emerging magnetic fields and observations, as shown in Figure~\ref{fig:sigmoids}. In the left hand column, isosurfaces of
current density show the evolution of sigmoidal structures, from their first appearance as two separate J-like structures that coalesce to form a single sigmoid in the middle row before starting to fragment at the end of the simulation. The regions of strong current density are the locations where Ohmic heating is likely to occur. In the middle column, the soft X-ray  plasma emission is estimated as $I = \int \rho^2 dz$, where the integration is along the line of 
sight and of plasma within the specific temperature range of the telescope bandpass. The final column presents the observations of a sigmoid from the X-Ray Telescope (XRT) on the Hinode satellite. Note that in the top row, the sigmoid does appear as two separate J-like structures. In the middle, the J-like structures have combined to form a single structure. In the bottom row, the observations show a very bright core at the centre of the sigmoid. This bright core appears to correspond to the formation of a new flux rope, as discussed in Section~\ref{subsec:rope}. The
reconnection responsible for the formation of the new rope heats the plasma locally, producing hot dense plasma, which is seen in soft X-rays. This flux rope can erupt, as discussed in Section~\ref{subsec:erupt} below, and these observations actually do show a dynamic bright feature moving away from the sigmoid in exactly the manner predicted by the simulations.

\subsubsection{Eruptions - the role of the overlying field}\label{subsec:erupt}
Observations of newly emerging magnetic regions frequently have associated eruptions of plasma that are sufficiently energetic to escape into interplanetary space. As mentioned in Section~\ref{subsec:rope}, many simulations \citep[e.g.][]{Manchester2004, Archontis2008} have found that a flux rope forms as part of the emergence process. These flux tubes rise but do not always erupt. So why do some fail to erupt while others do? 

Consider, first, a corona with no pre-existing magnetic field. As mentioned above, the first field to emerge from the interior is in a North-South direction. This field is strongly anchored in the dense photosphere. When the flux rope finally forms, by the process described above, it is not so securely held down because its footpoints are in the more distant sunspots and not in the nearby photosphere. It tries to breaks away from the photosphere but there is a significant amount of already emerged magnetic field above it, which is strongly anchored to the photosphere. The magnetic tension in this North-South field can prevent the full eruption of the flux rope. 
\begin{figure}
\centering
\includegraphics[width=0.7\textwidth]{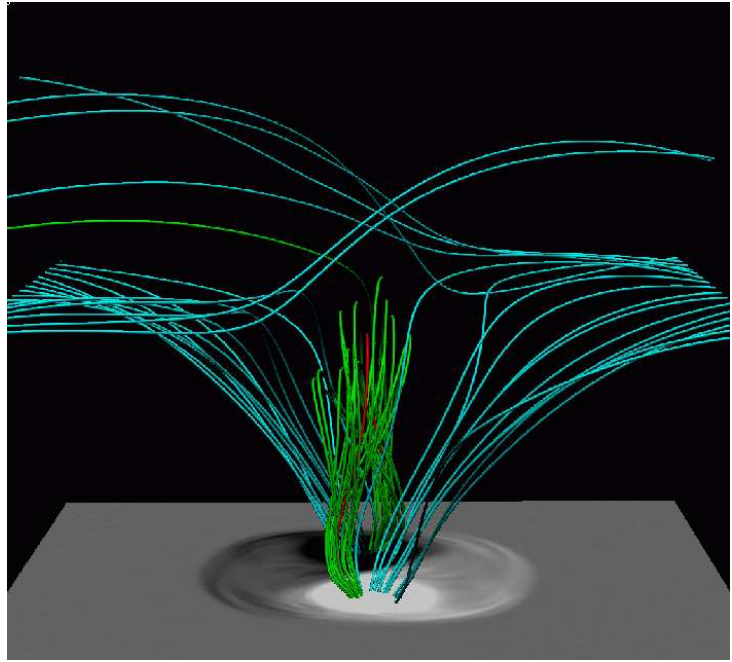}
\caption{Interaction with overlying field. \citep[From][]{MacTaggart2009}}
\label{fig:coronalB}
\end{figure}
Now consider emergence when there is a pre-existing coronal magnetic field and, in particular, one that is not aligned with the direction of the first field to emerge. Magnetic reconnection between these fields can remove the magnetic tension and allow the flux rope to fully erupt. 
Before the magnetic field emerges, there is an initially horizontal coronal magnetic field. As the new field emerges through the photospheric boundary, this rising field reconnects with the overlying field. Figure~\ref{fig:coronalB} shows how the reconnected coronal magnetic field lines (traced from the sides of the computational box) now drop down into the outer regions of the emerging flux tube. The newly formed flux rope is shown by green field lines. Since all the field above the flux rope has reconnected with the overlying field, there is no longer any magnetic tension to hold it down and the flux rope can erupt unhindered into an almost field-free region.

\section{Summary}
The recent theoretical ideas to explain the solar dynamo, covering both the generation of the large-scale and small-scale magnetic fields, have been presented. Small-scale magnetic dynamo action results in any suitably turbulent flow provided the magnetic Reynolds number is sufficiently high. Purely on theoretical grounds, therefore, it is to be expected that the granules and supergranules, confined close to the surface, will, of themselves, act to generate small-scale magnetic field. This idea is confirmed by observations that show that the amount of small-scale field is independent of the solar cycle, and hence is not simply a by-product of the large-scale dynamo.

Understanding the generation of the Sun's large-scale magnetic field ---  the field of the solar cycle --- remains one of the great challenges of astrophysical MHD. Although most dynamo modelling has been performed within the framework of mean field electrodynamics, there are significant problems in applying this theory to the astrophysically relevant, high $Rm$ regime. Current solar dynamo models fall into one of three broad categories: distributed dynamos, in which dynamo action takes place throughout the convection zone; interface dynamos, in which dynamo action is localised in the tachocline and lower convection zone; and flux transport dynamos, in which the two key components of the dynamo cycle (`$\alpha$' and `$\omega$') are separated by the extent of the convection zone but are connected by a large-scale meridional cell. As discussed in Section~\ref{Subsec:solar_dynamos}, there are problems with all three models, though the idea that the solar magnetic field is generated by a flux transport dynamo does seem the least plausible. Progress with the solar dynamo problem will result from a combination of high resolution numerical simulations, both global and local, focusing on the tachocline for example, together with an improved theoretical understanding of the fundamental aspects of MHD turbulence.

The eleven year sunspot cycle has a profound effect on the evolution of the global coronal magnetic field. For example, the amount of open magnetic flux tends to peak two years after solar maximum. This is due to the non-potential nature of the coronal magnetic field that inflates the corona, causing the opening of closed field lines. Properties of solar prominences can be difficult to predict but, by following the motions of the observed surface magnetic fields and, hence, the build-up of stresses in the coronal magnetic field, it is now possible to predict where prominences will form and which are likely to erupt as a CME. The ideas behind the global 
coronal field models, namely the inclusion of new emerging magnetic fields, the stressing due to shearing by differential rotation, the transport of magnetic flux to the poles, the break up of strong field regions and the decay of active regions through diffusion, can be used to follow the evolution of an individual active region. The time resolution of the order of 30 seconds, full disk observations
from the Solar Dynamics Observatory now mean that the line-of-sight magnetic field and the observed photospheric flow patterns can be used as input for this model and the subsequent evolution of the coronal magnetic field determined. Initially, the models will require the real-time 
surface flow velocity as input and the evolution of the resulting coronal magnetic field will be compared with observations. Next, more generic flow fields can be used to see if a similar evolution is generated and if similar dynamic features are seen. Then, in the future development of this model, it will be possible to predict whether a newly emerging sunspot or active region is likely to become unstable and generate a large solar flare or not.

One by-product of the large-scale dynamo is sunspots. They emerge, evolve and disperse on the order of a few weeks and the interaction between the emergence of these strong magnetic fields and the pre-existing coronal magnetic field creates a variety of dynamic solar phenomena. At present, numerical simulations can identify the physical processes responsible for the
emergence. However, current computational resources, mainly computer memory, mean that only the uppermost part of the solar convection zone and the low corona can be modelled. In addition, the simulation timescales are based on resolving the shortest timescales, determined by the propagation of Alfv\'en waves in the solar corona, and these are significantly shorter than the observed times for flux emergence. Hence, to save on computing time, the magnetic buoyancy of the emerging flux tubes is chosen to produce a faster evolution. Thus, the magnetic field strength at the photosphere increases much faster than would be liked and this means that the emergence process is also faster. Now this mismatch in timescales is not a major issue since the correct ordering of the timescales for each physical process is preserved, i.e.\ Alfv\'en wave timescale is shortest and so on. Nonetheless, the issue of timescales will be addressed with the next generation of parallel computers.

The simulations of flux emergence are now being performed with more realistic pre-existing magnetic environments. For example, as discussed in Section~\ref{subsec:rope}, the formation of a new flux rope, during the simulations of the emergence of magnetic fields, often results in the eruption of dense plasma from the lower corona. How closely are these eruptions related to the observed CMEs? Again the timescales are wrong but it is possible that the basic physical processes involving the removal of the overlying magnetic field (see Section~\ref{subsec:erupt}) are correct.

Another coronal feature frequently observed is the sigmoid, observed in X-rays. Sigmoids naturally arise in simulations 
from the emergence of a twisted flux rope. However, some of them seem to occur during the decaying phase of an active region. Nonetheless, although in this case new magnetic flux is not emerging, there does seem to be a shearing motion along the magnetic polarity inversion line and a general increase in the magnetic helicity. Both properties occur naturally in the emergence simulations. So again, the key physical processes are correctly described even when sigmoids do not exhibit an increase in photospheric magnetic flux.

We have described some of the wide variety of problems currently of importance in the 
generation and evolution of solar magnetic fields. The nature of the generation of the large-scale solar magnetic field is still far from understood, and remains one of the great challenges of astrophysical MHD. The continual improvement in the quality of the observations of the solar surface and atmosphere is helping to improve modelling of the photospheric and coronal fields; with the advances in computational resources, the simulations are now allowing us to compare theory and observations at a level never before achievable.

\section*{Acknowledgements}
We would like to thank the organisers, particularly Prof. Keke Zhang, for the invitation to attend BEPIS 2010 and for their hospitality at the meeting. Some of the computational work was performed on the UKMHD Consortium parallel computer that was jointly funded by STFC and SRIF.

\end{document}